\documentclass[aps,amssymb,amsfonts,amsmath,showpacs,twocolumn]{revtex4}

\usepackage{hyperref}

\usepackage{graphicx}
\usepackage{dcolumn}
\usepackage{bm}

\begin{document}

\title{Coupling of plasmonic nanoparticles to their environments in the context of van der Waals-Casimir interactions}

\author{U. H\aa{}kanson}
\altaffiliation{Present address: Division of Solid State Physics,
Lund University, Box 118, S-221 00 Lund, Sweden}

\author{M. Agio}
\affiliation{Laboratory of Physical Chemistry, ETH Zurich, 8093
Zurich, Switzerland}

\author{S. K\"{u}hn}
\altaffiliation{Present address: School of Engineering, University
of California, Santa Cruz, CA }

\author{L. Rogobete}
\affiliation{Laboratory of Physical Chemistry, ETH Zurich, 8093
Zurich, Switzerland}

\author{T. Kalkbrenner}
\altaffiliation{Present address: CyBio AG, Goeschwitzer Str. 40,
D-07745 Jena, Germany}

\author{V. Sandoghdar}
\email{vahid.sandoghdar@ethz.ch} \affiliation{Laboratory of Physical
Chemistry, ETH Zurich, 8093 Zurich, Switzerland}


\begin{abstract}
We present experiments in which the interaction of a single gold
nanoparticle with glass substrates or with another gold particle can
be tuned by \emph{in-situ} control of their separations using
scanning probe technology. We record the plasmon resonances of the
coupled systems as a function of the polarization of the incident
field and the particle position. The distinct spectral changes of
the scattered light from the particle pair are in good agreement
with the outcome of finite difference time-domain (FDTD)
calculations. We believe our experimental technique holds promise
for the investigation of the van der Waals-Casimir type interactions
between nanoscopic neutral bodies.
\end{abstract}


\maketitle

\section{Introduction}
The recent progress in nanotechnology has introduced an immediate
need for the optimal design of miniaturized devices and therefore a
better understanding and exploitation of the interactions between
nanoparticles and surfaces~\cite{Capasso:07}. In some applications
such as mechanical actuation, the main concern is about forces among
the nanoscopic components of the system~\cite{Chan:01}. In some
other cases, for example in optical sensing and imaging, the
radiation pattern, absorption, emission or scattering spectra of
nano-objects are at the center of
attention~\cite{Fritzsche:03,Lakowicz:06}. For a number of reasons,
metallic nanostructures have played an increasingly important role
in these developments. Aside from their electric conductivity,
nanoparticles made of noble metals can be chemically and
photophysically very stable. Moreover, they provide interesting
optical features such as spectral selectivity and the possibility of
enhancing optical fields via their plasmon resonances. In this
article, we present experimental results on the controlled study of
the optical interaction between a single gold nanoparticle and
dielectric surfaces as well as the coupling between two individual
gold nanoparticles. We discuss our measurements in the context of
the van der Waals-Casimir (vdW-C) interactions and point out the
potential of our experimental system for investigating the interplay
between the classical and quantum mechanical phenomena related to
vdW-C interactions.

\section{A gold nanoparticle as a model dipolar antenna}

\begin{figure}
\includegraphics[width=8.3cm]{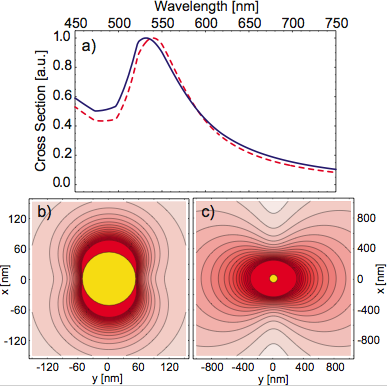}
\caption{\label{plspectr}(a) Plasmon spectrum of a gold nanoparticle
($D=100$~nm) in air ($\epsilon_\mathrm{m}=1$) computed using Mie
theory (blue line) and the effective polarizability of
Eq.~(\ref{alpha-eff}) (red dashed line). (b) and (c) represent the
amplitude of the scattered electric field calculated using Mie
theory in the near and far fields, respectively. The illumination
was a plane wave propagating along \emph{z} and polarized along the
\emph{x}-direction. The yellow disks depict the nanoparticle to
scale.} \label{dipolar-particle}
\end{figure}

The scattering properties of nanoparticles can be calculated in
terms of multipoles by using the theory developed by G.
Mie~\cite{Mie:08}. If a spherical particle has a diameter $D$ that
is much smaller than the wavelength $\lambda_m$ of the incident
light in the medium surrounding the particle, it can be treated as a
dipole with a polarizability~\cite{Bohren-83book}
\begin{equation}
\alpha (\lambda )=\frac{\pi D^{3}}{2}\frac{\epsilon _{p}(\lambda
)-\epsilon _{m}}{\epsilon _{p}(\lambda )+2\epsilon _{m}}
\label{alpha-sphere}
\end{equation} where $\epsilon _{p}(\lambda )$ and $\epsilon _{m}$ are the dielectric constants of
the particle and its surrounding medium respectively, and $\lambda$
is the vacuum wavelength. If the material conditions are such that
the denominator has a minimum at a certain wavelength, the
polarizability and therefore the scattering cross section are
enhanced. This can take place for metallic nanoparticles, leading to
localized plasmon resonances~\cite{Bohren-83book,Kreibig}. The
scattering cross section of a subwavelength particle is given by
\begin{equation}\sigma=\frac{(2\pi)^3 \epsilon_\mathrm{m}^2
|\alpha|^2} {3\lambda^4} \label{scs}.
\end{equation} It turns out that a gold nanoparticle placed in air or in a
low index dielectric material such as glass shows a resonance in the
visible domain (see Fig.~\ref{dipolar-particle}a). However, plasmon
spectra can be varied by changing the size, shape and the dielectric
functions of the medium surrounding the
particle~\cite{Bohren-83book,Kreibig}.

Gold nanospheres of diameter 50 to about 100~nm are in the regime
where the dynamical effects become important and the observed
plasmon resonance begins to deviate from that predicted by the
simple formula in Eq.~(\ref{alpha-sphere}). Nevertheless, it has
been shown that the plasmon spectra of such particles can be
reproduced very well if one takes into account radiation damping and
dynamic depolarization to arrive at an effective dipolar
polarizability given by~\cite{Meier:83}
\begin{equation}
\label{alpha-eff} \alpha_\mathrm{eff}(\lambda)=
\frac{\alpha(\lambda)\left(1-\frac{\displaystyle
\pi^2\epsilon_\mathrm{m}D^2}{\displaystyle 10\lambda^2}\right)}
{1-\frac{\displaystyle 2\pi\epsilon_\mathrm{m}} {\displaystyle
D\lambda^2}\alpha(\lambda)- i\frac{\displaystyle
4\pi^2\epsilon_\mathrm{m}^{3/2}} {\displaystyle
3\lambda^3}\alpha(\lambda)}.
\end{equation} Figure~\ref{dipolar-particle}a shows that the far-field scattering cross
section calculated according to Mie theory (blue) agrees quite well
with that evaluated by using $\alpha_\mathrm{eff}$ (dashed red). The
contribution of higher multipoles is negligible in the far field,
but it could amount to up to 10\% of the dipolar one in the near
field. This deviation becomes more important for larger particles or
higher values of $\epsilon_\mathrm{m}$.

It follows that the electric field lines of the light scattered by
the nanoparticle trace a dipolar radiation pattern.
Figures~\ref{dipolar-particle}b and c display the strength of the
scattered electric field in the near and far fields of the particle,
respectively. Thus, gold nanospheres smaller than about 100~nm in
diameter can be treated as subwavelength classical dipolar
antennae~\cite{Kalkbrenner:05,Kuehn:06}.

The dipolar character of a gold nanoparticle makes it an ideal
approximation to a point-like oscillator, which has been a very
useful conceptual construct for relating the classical and quantum
mechanical features of an atom~\cite{Loudon,Haroche-Houches}. Here,
the plasmon resonance plays the roles of the transition between the
ground and the excited states. The Rayleigh-like scattering replaces
spontaneous emission~\cite{Mazzei:07} whereby the particle
polarizability and hence its scattering cross section $\sigma$
provide a measure for the strength of this process. Furthermore,
absorption in a gold particle mimics a nonradiative decay channel in
an atom~\cite{Buchler:05}. Thus, one could expect that optical
interactions that influence the spectrum of an atom would have an
analog in the context of plasmon resonances of small gold particles.

Of particular interest to the topic of this article is that the
plasmon spectrum of a gold nanoparticle can be broadened in a medium
of higher dielectric constant~\cite{Kreibig}, similar to the change
of the atomic fluorescence decay rate in a dielectric
medium~\cite{nienhuis01}. Interestingly, a quick examination of the
expression in Eq.~(\ref{scs}) shows that as in the case of the
Einstein A coefficient, the rate of scattering obtained according to
$I_{inc}\sigma/\hbar \omega$ is proportional to $\omega^3$ and to
the square of the dipole moment via $\alpha^2$. In fact, we recently
showed in Ref.~\onlinecite{Mazzei:07} that Rayleigh scattering can
be also modified if the density of states are manipulated, for
example by a high-Q cavity. To this end, one could expect that the
plasmon spectrum be modified close to an interface, as was discussed
by A. Sommerfeld for a classical dipolar
antenna~\cite{Sommerfeld:09,Morawitz:69,Barnes:98} and by F. London
for a quantum mechanical atom~\cite{London:30b}. In the latter case,
the fluctuating quantum mechanical dipole moment of the atomic
ground state is assumed to undergo an instantaneous interaction with
its image in the nearby mirror. London also showed that the origin
of the van der Waals interaction between two atoms is their
dipole-dipole interaction~\cite{London:30a}. Thus, similarly there
should be a corresponding van der Waals interaction between a pair
of gold nanoparticles.

In this article, we explore such effects by studying the
modifications of the plasmon resonances from nanoparticles under
controlled conditions. A systematic experimental investigation of
the interaction between an oscillating dipole and its surrounding
requires a change of its position with an accuracy well beyond
$\lambda/2\pi$ where $\lambda$ is the dipole oscillation wavelength.
This has been explored in atomic physics
experiments\cite{Sandoghdar:92,Fichet:95,Sandoghdar:96}, but neither
at the single atom level, nor with nanometer precision. Here we show
that both of these issues can be addressed using nanoparticles in
the condensed phase.

\section{Experimental approach}\label{experimental}
\begin{figure} \centering{
\includegraphics[width=8 cm]{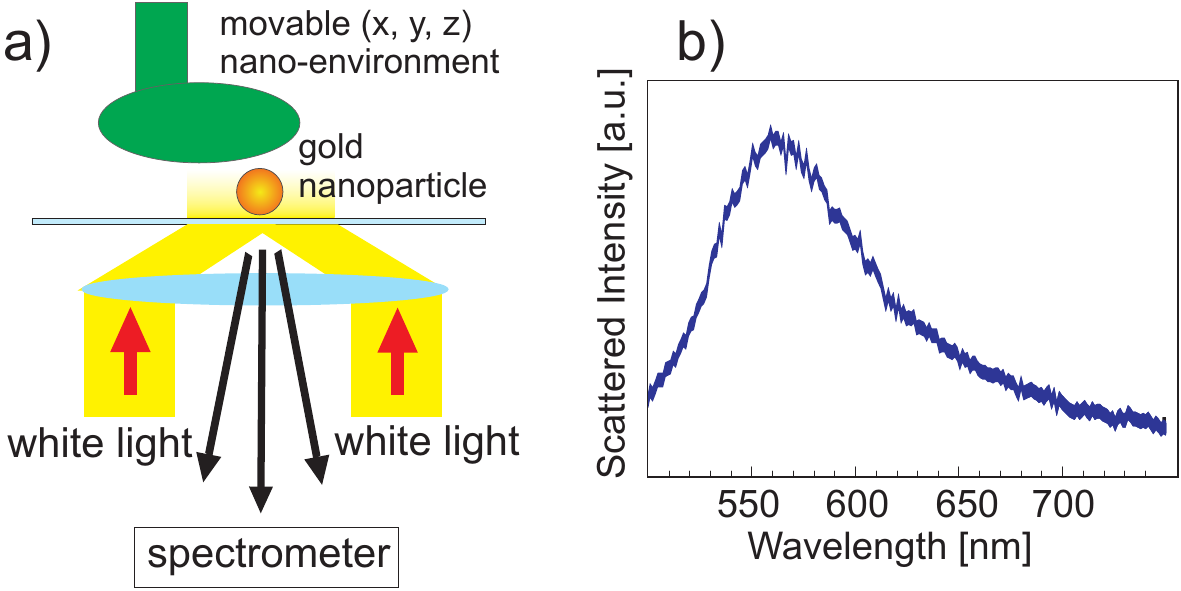}}
\caption{a) White light from a Xe lamp is illuminated onto a gold
nanoparticle in the total internal reflection mode. The scattered
light from the particle is collected and sent to a spectrometer. A
scanning probe stage can move a second object in the immediate
vicinity of the particle. b) A typical scattering spectrum collected
from a single gold nanoparticle on a glass substrate. }
\label{setup-resonance}
\end{figure}

Figure~\ref{setup-resonance}a displays the basic setup for recording
the plasmon spectrum using an inverted optical microscope. A white
light source is used to illuminate the sample in total internal
reflection using an immersion oil microscope objective. By proper
masking of the incident white light beam~\cite{Gunnarsson:05} and
using a polarizer, we select a well-defined direction and
polarization of illumination. The scattering from the particle is
collected by the same microscope objective, passed through a
confocal pinhole (not shown) and directed to a spectrometer.
Figure~\ref{setup-resonance}b shows an experimentally recorded
plasmon spectrum of a single gold nanoparticle of diameter 100~nm on
a glass substrate.

A home-built scanning near-field optical microscope (SNOM) and a
piezo-electric stage are used to move a second object attached to a
fiber tip in the immediate vicinity of a gold nanoparticle. The
shear-force mechanism~\cite{Karrai:00} allows us to control the
separation between this object and the gold particle. To do this,
the tip is vibrated parallel to the substrate. When the tip comes
close to the surface, it experiences a lateral force that dampens
its oscillation amplitude. By monitoring this change, the tip can be
kept at a constant gap from other objects. Depending on the material
of the tip and the substrate, the tip geometry, and the applied
oscillation amplitude, the range of the shear-force interaction can
be different from a few to a few tens of
nanometers~\cite{Karrai:00}. More details about the setup can be
found in Refs.~\onlinecite{Kalkbrenner:01,Kalkbrenner:05,Buchler:05}

\section{The interaction between a gold nanoparticle and two dielectric mirrors}

\begin{figure} \centering{
\includegraphics[width=7 cm]{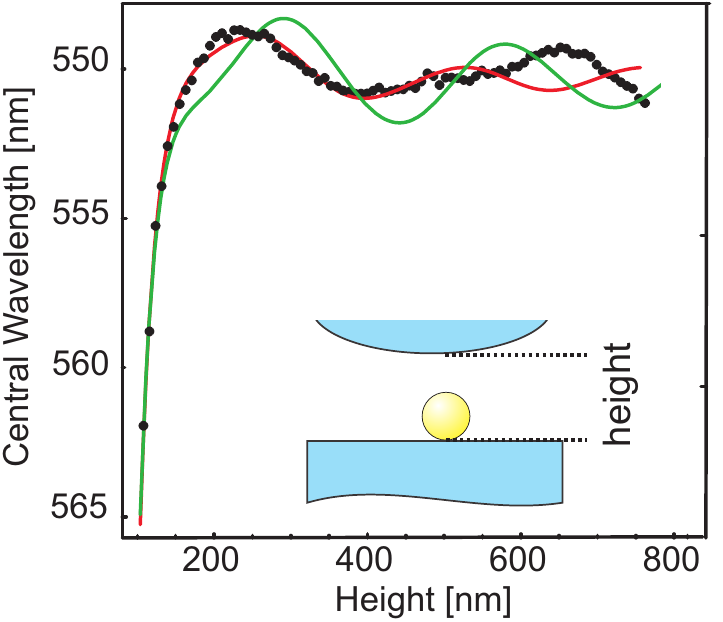}}
\caption{The peak wavelength of the plasmon resonance of a gold
nanoparticle placed on a glass substrate as a function of the
displacement of a second slightly curved glass surface, as sketched
in the inset. The green curve shows the prediction of an analytical
calculation for a point-dipole. The red curve takes into account an
interference artefact discussed in the text.} \label{gnp-glass}
\end{figure}

In this section, we examine the interaction of a single gold
nanoparticle placed on a glass substrate with a movable glass
mirror, as depicted in the inset of Fig.~\ref{gnp-glass}. To prepare
the sample, we spin coated gold nanoparticles of diameter 100~nm
(British Biocell International) on a glass cover slide with a dilute
coverage of the order of one particle per $10~\mu m^2$. The incident
light could be adjusted to have a $p$ or $s$ polarization. A pinhole
is placed in the detection path in order to reduce the background
scattering and collect the signal from a very small region of the
order of one micrometer. For the movable mirror, we used a
microsphere of diameter $60~\mu m$ melted at the end of an optical
fiber\cite{Buchler:05}. The fiber was mounted in the SNOM stage so
that the microsphere surface could be approached to the gold
nanoparticle with nanometer precision. Given that we investigate
very small separations of the order of one micrometer, in what
follows we approximate this slightly curved surface by a flat
mirror.

Determining a shift or broadening of the plasmon resonance can be a
subtle task because the plasmon resonance is not symmetric and does
not possess a simple line profile. To assign the resonance peak, we
fitted each plasmon spectrum by using the dipolar approximation and
an effective polarizability of Eq.~(\ref{alpha-eff}), as described
in Ref.~\onlinecite{Buchler:05}. Alternatively, we used the
procedure in Ref.~\onlinecite{Kuwata:03}. Using both methods, we
could determine the peak of the resonance with a resolution of about
0.5~nm. Next, by comparing the plasmon spectrum recorded in the
presence of the movable curved glass surface with that in its
absence, a spectral shift was extracted for each position. The
symbols in Fig.~\ref{gnp-glass} show the shift of the plasmon
resonance recorded at different surface-surface separations for
\emph{p}-polarized illumination. We see clearly that the plasmon
resonance experiences a red shift with a steep gradient at close
distances. This can be understood as the attractive near-field
interaction of the plasmon oscillation and its instantaneous mirror
image in the nearby surface. Such an energy shift has been
previously reported for ensembles of metallic
nanostructure~\cite{Holland:84,Murray:06}, but to our knowledge this
is the first report on the spectral shift of single particles.

At larger separations, Fig.~\ref{gnp-glass} reveals oscillations,
again as expected from the theoretical work on the interaction
between a classical dipole and a mirror~\cite{chance:78,Hinds:91}.
These oscillations have also been reported recently for ensembles of
metallic nanostructures~\cite{Murray:06}, but their analysis
requires a great deal of care. The oscillations seem to be irregular
and deviate from the theoretical values (green curve) obtained by
considering the frequency shift of a classical dipole placed between
two flat glass surfaces~\cite{Sullivan:97}. In these calculations a
point-like dipole was fixed at a distance of 50~nm (the radius of
the gold nanoparticle) from a glass substrate and a second glass
surface was scanned. We believe the cause of these deviations is a
subtle interference effect between the light scattered by the
particle and other objects (e.g. the substrate) in its vicinity. In
the particular case of the current experiment, the evanescent
illumination can be scattered by the gold particle and subsequently
reflected back into the detection path by the second curved surface
in the vicinity. Since in these experiments the separations between
the objects under consideration are very small, the background
scattering is coherent with the light that is scattered from the
nanoparticle of interest. It turns out that despite the use of a
pinhole in detection, some of the background scattering reaches the
detector and interferes with the field from the particle. As
distances are changed, this interference gives rise to
wavelength-dependent intensity modulations and skews the broad
plasmon resonance by a small amount, thus causing a systematic error
in the extracted resonance wavelength peak values. We have verified
that this effect is stronger for the s-polarized illumination
because this light is more efficiently scattered along the axis of
the microscope objective. A detailed account of this effect can be
found in the supplementary materials of
Ref.~\onlinecite{Kalkbrenner:05}. Our efforts to fully eliminate
this systematic effect have not been successful so far. However,
this phenomenon can be crudely taken into account in the fit to the
experimental data by allowing for a small sinusoidal function added
to the theoretically calculated values. The outcome of such a fit is
shown by the red curve in Fig.~\ref{gnp-glass}. The fact that the
correction is no longer effective for larger distances might
indicate that the background scattering stems from more than one
point. We emphasize, however, that this systematic problem is not of
concern for near-field measurements.

We note that the point-dipole model breaks down when a nanoparticle
is closer than about a particle diameter. Nevertheless, the orders
of magnitude and the general trend of the calculated modification of
the plasmon resonance seems to yield a good agreement with the
experimental results discussed above and in
Refs.~\onlinecite{Kalkbrenner:05,Buchler:05}. In the next section,
we discuss the interaction between two nanoparticles and compare our
experimental data with the outcome of numerical simulations.

\section{The interaction of two gold nanoparticles}
In order to investigate the coupling of two gold nanoparticles in a
well-defined manner, we have exploited scanning probe technology to
control their relative positions \emph{in-situ}. One gold particle
($P_\mathrm{tip}$) is attached to the end of a glass fiber tip,
following the preparation technique that we have reported
previously~\cite{Kalkbrenner:01,Kalkbrenner:04}. The second particle
($P_\mathrm{sub}$) is selected from a very low concentration of
nanoparticles (diameter of 100~nm) spin coated on a glass substrate.
The plasmon resonances of each particle can be recorded separately
in the configuration shown in Fig.~\ref{setup-resonance}a.

\begin{figure}
\centering{
\includegraphics[width=8 cm] {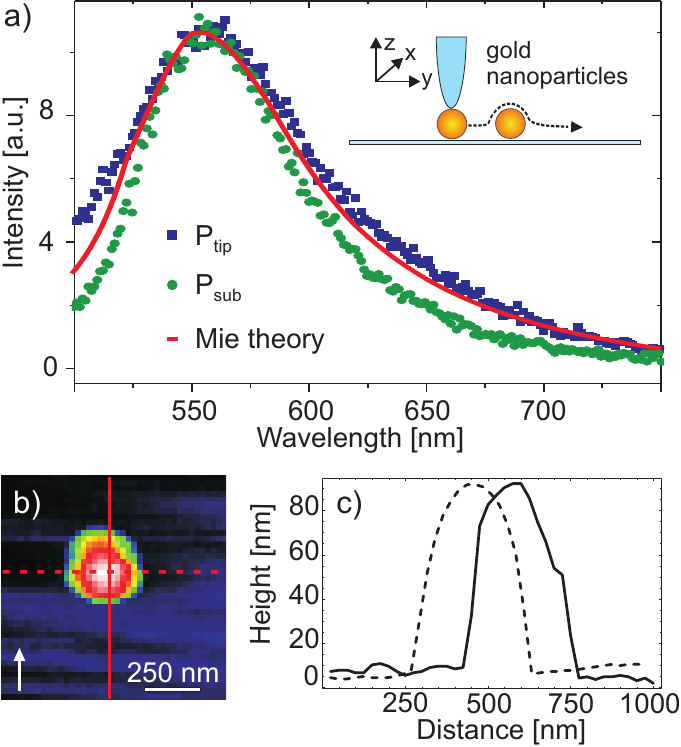}}
\caption{(a) Measured scattering signal from the 100~nm gold
particles on the tip (blue symbols) and on the substrate (green),
respectively. The red curve shows the calculated spectrum for a
100~nm gold particle using Mie theory. Inset: schematics of the
experimental arrangement. (b) Topographic image obtained when
scanning the single particle probe across the particle fixed on the
substrate. The white arrow indicates the electric field direction of
the excitation light. Note that the particle appears larger because
of the convolution with the finite size of the tip. (c) Cross
sections along the solid and dashed lines in (b).}
\label{schematics}
\end{figure}

The blue and green symbols in Fig.~\ref{schematics}a show the
plasmon spectra of $P_\mathrm{tip}$ and $P_\mathrm{sub}$ recorded
independently, under \emph{s}-polarized incident light. The two
particles were chosen to have very similar plasmon spectra.
Furthermore, by examining these spectra as a function of the
polarization of the excitation light, we verified that they were
spherical~\cite{Kalkbrenner:04}. The red trace in
Fig.~\ref{schematics}a represents a calculated Mie spectrum for a
100~nm spherical particle using the tabulated dielectric function of
gold~\cite{JohnsonChristy:72}. In order to obtain the best agreement
with the spectra of both $P_\mathrm{tip}$ and $P_\mathrm{sub}$, an
effective refractive index of $1.18$ for the surrounding medium was
taken. Here, one has to bear in mind that the immediate environments
of $P_\mathrm{tip}$ and $P_\mathrm{sub}$ (i.e. a glass tip and a
glass substrate) are asymmetric, so that it is a nontrivial matter
to define a dielectric function for the medium surrounding the
nanoparticles. Nevertheless, it has been previously shown that a
weighted average of the dielectric constants of the various
surrounding media yields reasonably good agreement with experimental
data~\cite{Tamaru:02,Hilger:01}. In what follows, the red curve in
Fig.~\ref{schematics}a will be used as the basis for the theoretical
modelling of the particle-particle interaction.

\begin{figure}
\centering{
\includegraphics{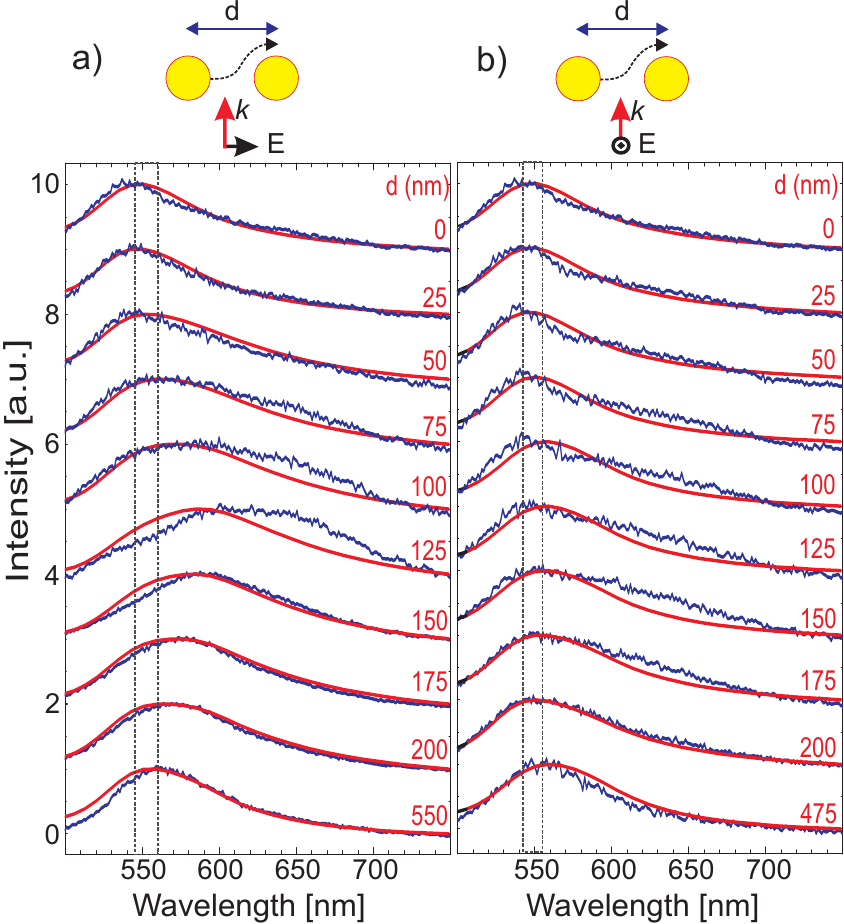}}
\caption{Series of spectra acquired for the particle pair with
polarization of the exciting field in (a) parallel to the particle
pair axis and in (b) orthogonal to the particle pair axis. The
spectra have been normalized and shifted in intensity for clarity.
Calculated scattering spectra for the situations shown as red lines.
The paremeter $d$ denotes the projection of the center-to-center
particle separation onto the substrate plane. The dashed vertical
lines mark the plasmon resonance peaks for \emph{d} large and
$d=0$.} \label{spectra}
\end{figure}

After determining the optical properties of the two particles
individually, we set out to study the interaction between them. The
drawing in Fig.~\ref{schematics}a shows the schematics of the
experimental arrangement. Using the SNOM stage, we can position both
the tip and the sample with nanometer accuracy. Furthermore, shear
force tip-sample distance control allows for combined topographic
and optical studies. Figure~\ref{schematics}b shows the measured
topography of $P_\mathrm{sub}$ as $P_\mathrm{tip}$ was scanned
across it, and Fig.~\ref{schematics}c displays the cross sections
corresponding to the two cuts parallel and perpendicular to the
polarization of the illumination. At each scan pixel of 25~nm, we
trigger the spectrometer and record the scattering spectra of the
two gold nanoparticles. Figures~\ref{spectra}a and b show selections
of spectra from the first halves of the scans marked by cross
sections shown in Fig.~\ref{schematics}b. In an intuitive picture
where each particle acts as an induced electric dipole moment
directed along the incident field polarization, these two series
explore the dipole-dipole interaction for the head-to-tail and
side-by-side configurations,
respectively~\cite{Quinten:98,Rechberger:03,Dahmen:07}.

Let us first concentrate on the case of Fig.~\ref{spectra}a. As the
two particles are brought closer from the large separation of
$d=550$~nm, the pair spectrum becomes broader and is shifted to the
longer weavelength. Here the parameter $d$ denotes the projection of
the center-to-center particle separation onto the substrate. This
was extracted from the topographic information recorded
simultaneously to the optical spectra, taking into account a
vertical offset of 15~nm due to the shear force interaction length
between tip and substrate~\cite{Karrai:00}. At $d \sim 125 $nm the
spectral shift amounts to a maximum of about $60$~nm. As the tip is
scanned further in shear-force control, $P_\mathrm{tip}$ is lifted
upward, resulting in a narrower and blue-shifted spectrum. When
$P_\mathrm{tip}$ is above $P_\mathrm{sub}$ (i.e. $d=0$), the
spectrum has become even \emph{narrower} than the starting point,
and its center wavelength is shifted to a lower wavelength by about
15~nm.

To compare our findings with theoretical expectations, we have
performed Finite-Difference Time-Domain (FD-TD) calculations of the
total scattering cross section of two spherical particles with
diameters of 100~nm~\cite{Taflove-Hagness,Oubre:04}. We treat the
dielectric function of gold using a Drude-Lorentz
model~\cite{Vial:05}. The space discretization has a step size of
$2$~nm, which has been shown to be small enough to obtain results
with $5\%$ accuracy.~\cite{Oubre:04} To ensure that the scattered
light exits the computational volume without reflections, we used
Convolutional-Perfectly-Matched-Layers (CPML) absorbing boundary
conditions~\cite{Roden:00}. The calculations took into account the
experimentally measured trajectory of the moving particle extracted
from the topography information (see Fig.~\ref{schematics}). For
positions where the tip is lifted from the substrate, the effective
refractive index of the surrounding media had to be reduced to $1.1$
(corresponding to a $6\%$ decrease) in order to obtain a good
agreement with the experimental results. We note that consideration
of an effective index for the glass-air interface is common also
when using fully analytical Mie theory~\cite{Kreibig}.

\begin{figure}
\centering{
\includegraphics[width=7 cm]{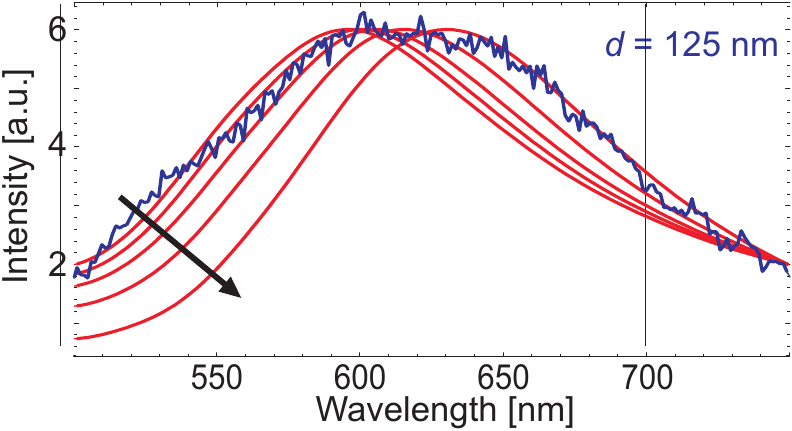}}
\caption{A comparison between experimental and calculated spectra
for close distances. The blue trace displays the measured scattering
signal for the particle pair with a nominal $d=125$~nm. Red curves
show the calculated spectra for distances from 130~nm to 110~nm in
steps of 5~nm (in direction of the arrow).} \label{inhomog}
\end{figure}

The agreement between the measured and the calculated spectra in
Fig.~\ref{spectra}a are generally very good aside from the region
about $ d \sim 125$~nm. At such small separations the plasmon
resonance is very sensitive to the exact distance between the
particles. To display this sensitivity to position, in
Fig.~\ref{inhomog} we compare the experimental spectra for
$d=125$~nm with five calculated plasmon spectra in which $d$ is
varied between $130$~nm and $110$~nm. We see that changing the
particle distance by merely $20$~nm shifts the peak of the plasmon
spectrum more than $40$~nm. Indeed, it is this sensitivity to
displacements that constitutes the core idea of the recently
proposed ``plasmon rulers."~\cite{Soennichsen:05,Reinhard:05} During
the analysis of these results, we have realized that the tip
oscillations (see section~\ref{experimental}) were excited too
strongly in these experiments. We have verified this hypothesis by a
direct measurement of the tip oscillation amplitude under the same
experimental conditions.

\begin{figure}
\centering{
\includegraphics{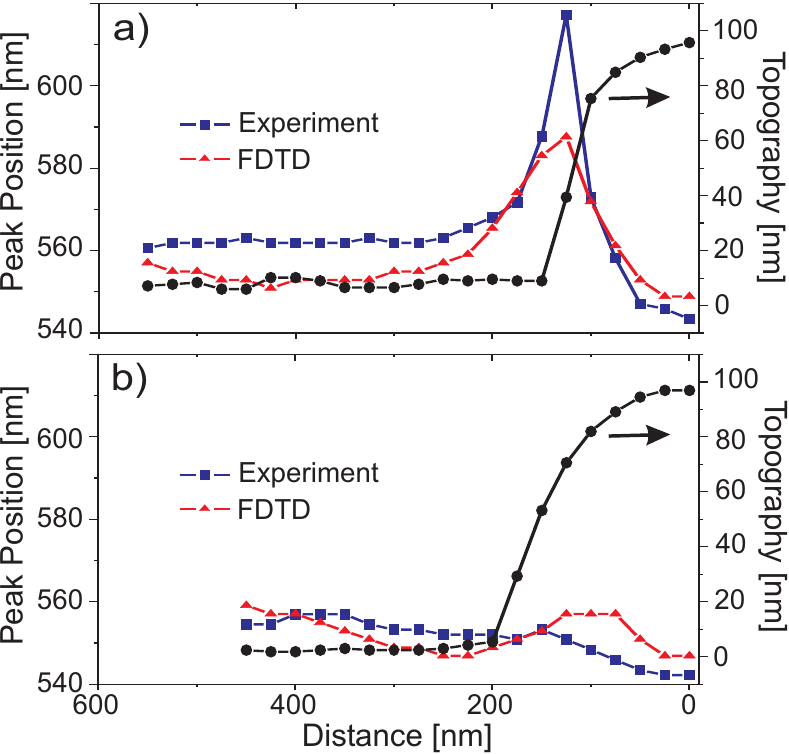}}
\caption{(a) Black circles display the cross section of the measured
topography for a scan along the lower half of the solid line in
Fig.~\ref{schematics}b. Blue squares and red triangles display
respectively the maximum peak position of the corresponding measured
and calculated spectra shown in \ref{spectra}a. In (b) the
topography and the peak positions correspond to a scan along the
left half of the dashed line in \ref{schematics}b. }
\label{spectra-summary}
\end{figure}

Figure~\ref{spectra-summary}a summarizes our results from the
configuration depicted in Fig.~\ref{spectra}a. The experimental
(blue) and theoretical (red) peak wavelengths are plotted as a
function of the tip position together with the corresponding
topography cross section. For large $d$, $P_\mathrm{tip}$ and
$P_\mathrm{sub}$ lie along the incident electric field. As $d$ is
decreased, we observe a shift of the resonance to the longer
wavelengths, corresponding to an attractive interaction expected for
the induced particle dipoles in a head-to-tail orientation. This red
shift is also accompanied by a broadening, which can be attributed
to a stronger radiative decay of the coupled system. Although the
system of two 100~nm particles no longer behaves as a pure dipolar
radiator, the simple picture of two dipoles adding up to yield one
larger dipole provides the correct intuition. As $P_\mathrm{tip}$
mounts $P_\mathrm{sub}$, the dipole arrangement quickly changes to a
side-by-side configuration, and the interaction becomes strongly
repulsive. For $d=0$, the plasmon resonance of the coupled system is
blue shifted by about 15~nm with respect to the non-interacting
spectra. All these trends are recovered by the FDTD calculations.
Moreover, the theoretical results predict a very small blue shift of
the spectrum in the region $500>d>300$ due to oscillation that stem
from the retarded interaction energy between two
dipoles~\cite{Dahmen:07}. Our experimental data do not show this
behavior. However, here one has to remember that the small
deviations between the plasmon resonances of the individual
particles (see Fig.~\ref{schematics}a) are not negligible at this
level of comparison between theory and experiment.

The situation for the configuration of Fig.~\ref{spectra}b is
presented in Fig.~\ref{spectra-summary}b and is quite different. In
comparison to the case of Fig.~\ref{spectra-summary}a, the
interaction is weaker. Here the polarization is perpendicular to the
axis joining the particles so that the relative orientation of the
dipoles in the two nanoparticles does not change during the entire
scan. As the particles approach each other, there is a shift towards
lower wavelengths, signifying a repulsive force. Again, the
theoretical calculations show the same trend. The difference between
the measured and theoretical peak positions at small distances stems
from a lack of perfect agreement between the recorded and calculated
lineshapes shown in Fig.~\ref{spectra}b.

Next, we remark on the widths of the plasmon resonances. Again,
aside from the separations around $ d \sim 125$~nm, the agreement
between the experimental and the FDTD results are quite good. We
find that in Fig.~\ref{spectra}a the attractive coupling between the
two particles is always accompanied by a spectral broadening. An
intuitive interpretation of this result is that at close distances
the head-to-tail arrangement gives rise to a total dipole moment
that is larger than that of each particle. As a result, the system
can radiate more strongly. The strong fields between the particles
in this arrangement are also responsible for the large enhancements
obtained in surface enhanced spectroscopy.~\cite{Moskovits:85}
Nevertheless, we emphasize that the simple picture of dipole-dipole
coupling breaks down when the particles are very close to each
other, and the contribution of multipoles should be taken into
account~\cite{Nordlander:04}.

By changing the interparticle distance and the orientation of the
pair relative to the polarization of the excitation, the plasmon
spectrum of the coupled system has been tuned by about 70~nm. On the
one hand, this shows the feasibility of the recent proposal for
using the distance dependent modification of plasmon resonances as a
measure for separation~\cite{Soennichsen:05}. On the other hand, the
data in Figs.~\ref{spectra} and \ref{spectra-summary} emphasize that
great care must be taken when extracting distance information
because changes of the polarization in relation to the particle pair
axis has a dramatic effect on the interaction. Moreover, variations
in the dielectric functions of the immediate surrounding could
strongly affect the scattering spectra.~\cite{Kalkbrenner:05}

\section{Discussion}

Modification of plasmon resonances of close lying particles have
been studied previously using various approaches. Most commonly,
researchers have examined ensembles of particles that have been
nanofabricated or chemically
produced~\cite{Rechberger:03,Su:03,Atay:04,Tamaru:02,Wei:04,Gunnarsson:05,Sundaramurthy:05}.
To remedy the inevitable inhomogeneities of prefabricated systems,
optical trapping of a single pair of silver
particles\cite{Prikulis:04} and directed assembly of silver and gold
dimers\cite{Soennichsen:05} have been also attempted. Our
experimental approach has the great advantage that a single
nanoparticle can be positioned with nanometer accuracy. In addition,
our system allows one to explore the analogy between the
modifications of plasmon resonances and vdW-C interactions.

Shortly after London's quantum mechanical description of the van der
Waals interactions~\cite{London:30a,London:30b}, Casimir and Polder
showed that a treatment based on quantized electromagnetic fields
introduces corrections to the London model~\cite{Casimir:48b}. In
the case of an atom in front of a mirror, the vacuum field
experiences boundary conditions and the local density of states are
altered, leading to the modification of the energy levels and of the
atomic fluorescence lifetime. The functional forms of the vdW-C
interactions turn out to be different for the ground and excited
states. At short atom-surface distances $d\ll\lambda$, the
interaction between the atom and the surface can be considered as
quasi-instantaneous so that both the ground and the excited states
experience a $1/d^3$ decrease in energy. At larger distances,
retardation takes effect and the ground state energy drops as
$1/d^4$ whereas the excited state undergoes oscillatory changes in
the energy and decay rate
\cite{Drexhage:74a,Haroche-Houches,Hinds:91,Barnes:98}. A similar
situation takes place if the interaction between two atoms is
considered~\cite{Casimir:48b}. In a physical picture, the atoms are
polarized by the fluctuating vacuum fields at all possible
frequencies, k-vectors and polarizations. When the separations are
much smaller than a transition wavelength, the induced dipole
moments in the atoms are quasi-instantaneous so that their
interaction can be treated as an electrostatic dipole-dipole
coupling, yielding a $1/d^6$ interaction energy~\cite{London:30a}.
For atoms placed at large separations, the finite speed of light
leads to an effective dephasing of the dipole moments, changing the
distance dependence of the energy from $1/d^6$ to $1/d^7$. The
above-mentioned interactions establish the fundament of the forces
between neutral bodies.

Interestingly, many of the features of the vdW-C interactions of
excited atoms can be described by considering an atom as a
``point-like" classical dipolar antenna that interacts with its own
radiated field after reflection from the
boundaries~\cite{chance:78}. However, the classical picture does not
match the quantum electrodynamic calculations fully. For example, it
has been shown that the excited state energy of the atom cannot be
described by a classical dipole near and far from the mirror
simultaneously~\cite{Meschede:90,Hinds:91}. This observation and the
fact that the stability and energy shift of the ground state cannot
be described by classical models have raised many interesting
discussions regarding the role of quantum electrodynamics in the
interaction of neutral bodies~\cite{Milonni-book}. It would be thus
intriguing to realize systems where the similarities and differences
between a classical point-like dipole and a quantum mechanical atom
can be studied experimentally.

Subwavelength gold nanoparticles might provide a promising system
for this purpose. On the one hand, such particles (diameter larger
than about 10~nm) are much larger than single atoms so that one can
use the dielectric function of bulk gold and classical
electromagnetic methods to describe their optical properties. In
this regard, they can be seen as the miniaturized versions of
neutral bodies that have been commonly used for studying vdW-C
forces~\cite{Lamoreaux:97,Mohideen:98,Roy:99}. On the other hand,
they can be considered to be point-like since they are much smaller
than a wavelength. The plasmon resonance of a gold nanoparticle lets
it act as a classical dipolar antenna whose radiation behaves like
that of an atomic excited state. At the same time, a nanoparticle is
similar to the ground state of a quantum mechanical atom in the
sense that it does not emit light and is radiatively stable.

Modification of the plasmon spectra of gold nanoparticles in the
near field of an interface has been verified
experimentally~\cite{Holland:84,Buchler:05,Kalkbrenner:05,Murray:06}
and has been described using classical models. Recently a few
publications have viewed such interactions in the light of
vdW-Casimir forces~\cite{Ford:98,Noguez:04,Noguez:04b}. Although the
plasmon spectra can be determined by solving classical Maxwell
equations, one can formulate these interactions also in the common
language of cavity quantum electrodynamics. In this picture, the
vacuum field excites plasmon oscillations in the gold nanoparticle
much as in the case of a ground state atom. If external boundary
conditions are present, the density of states are modified, leading
to the change of the plasmon spectrum. A similar point of view has
been also discussed regarding the role of the surface plasmons in
the attraction of metallic mirrors~\cite{Intravaia:05}. Thus, by
comparing the near- and far-field modifications of the plasmon
spectra and by treating the plasmon oscillations quantum
mechanically, one might be able to bridge the quantum world of atoms
and the classical realm of dipolar antennae.

Although it might be tempting to establish a direct link between the
plasmon resonance shifts summarized in Fig.~\ref{spectra-summary}
and the vdW-C interaction energies for the coupled particle system,
one has to remember that our measurements have recorded spectra for
particular directions of illumination and polarizations. However,
calculations of the vdW-C interaction energy should take into
account contributions of vacuum fluctuations at all wavelengths,
polarizations and k-vectors. Nevertheless, given that the
``response" of the system is the same for excitations by virtual and
real photons, it might be possible to extrapolate the laboratory
measurements to infer the vdW-C interaction energy. Such theoretical
considerations go beyond the scope of our current work, but very
recent advances in calculation of Casimir forces~\cite{Emig:07}
might be able to investigate such matters.

It is also instructive to consider the orders of magnitudes of the
forces involved. The largest measured gradient of the
energy-distance curves in Fig.~\ref{spectra-summary}a amounts to
$\Delta \lambda/\Delta z \sim 30~\mathrm{nm}/25~\mathrm{nm}$ which
corresponds to a force of $\sim 0.5$~pN. We estimate the smallest
force that is detectable with our signal-to-noise ratio to be a few
tens of fN, which is nearly two orders of magnitude smaller than the
sensitivity of the recent force
experiments~\cite{Mohideen:98,Chan:01}. Investigation of the vdW-C
interactions of metallic nanoparticles via spectroscopy could have
several advantages over direct force measurements. In particular,
one would be less sensitive to residual effects such as
electrostatic forces. Moreover, the optical constants of the
particles under study could be measured
\emph{in-situ}~\cite{Stoller:06}.

From an experimental point of view, the controlled laboratory
measurements of the vdW-C interactions and comparisons with theory
remain not only challenging for atoms and macroscopic objects, but
also for nanoparticles. First, the interactions are weak except at
very small separations. Second, they are strongly distance dependent
so that nanometer distance control is required. Third, the
geometrical and material features of the objects play an important
role and are not always well defined. In order to address these
experimental issues and realize a well-defined and controllable
system, we have followed a simple strategy to place a single
nanoparticle at the end of a sharp glass fiber tip and have used
scanning probe technology to place it in nanoscopic confined
geometries~\cite{Kalkbrenner:01,Buchler:05,Kalkbrenner:05,Kuehn:06}.
This approach has been very fruitful and has been also pursued by
other groups \cite{Eah:05,Anger:06,Danckwerts:07} for performing
controlled measurements with individual nanoparticles.

\section{Conclusion}
We have studied the weak interaction between a well characterized
single nanoparticle and a macroscopic dielectric mirror or a second
gold nanoparticle. Both red and blue shifts of the plasmon spectrum
as well as its broadening and narrowing were observed. Several
aspects of these experiments were not ideal and could be improved
for more precise measurements. In the first experiment, residual
reflections that can interfere with the light scattered by the
particles should be minimized~\cite{Kalkbrenner:05}. In the latter
experiment, more care should be given to maintaining a minimal
oscillation of the tip to avoid broadening of the spectra. It would
be particularly interesting to perform this experiment in immersion
oil to eliminate the influence of the glass substrate and the tip.

We have discussed our measurements in the context of the van der
Waals-Casimir interactions between neutral bodies. Both theoretical
and experimental reports in the literature have typically considered
two extreme cases of vdW-C interactions where the objects are either
single atoms and molecules or they are macroscopic bodies with
dimensions larger than a few micrometers. Scanning probe
manipulation and spectroscopy of very small gold nanoparticles could
allow access to an intermediate regime. Mounting a nanoparticle at
the end of an AFM tip would also provide the possibility of
simultaneously measuring the force and spectral modifications. In
particular, such spectroscopic studies could be easily extended to
metallic particles of the order of 5 nm in
diameter~\cite{Lindfors:04,Stoller:06}, which bridge the quantum
mechanical and classical regimes.

Finally, our experiments could be considered as the \emph{in-situ}
realization of a tunable optical
nanoantenna~\cite{Greffet:05,Muhlschlegel2005}. We have recently
shown that a pair of gold nanoparticles can act as a very efficient
antenna system for enhancing the spontaneous emission of an emitter
placed in their gap~\cite{Rogobete:07}. Given that the performance
of such devices requires an extremely high accuracy in the relative
coordinates and orientation of both the emitter and the
nanoparticles, a tunable setup~\cite{Kuehn:06} will be of great
advantage.

\end{document}